\documentclass[prl,twocolumn,superscriptaddress,showpacs,amsfonts,amsmath,floatfix]{revtex4}

\usepackage{graphicx}
\usepackage[usenames]{color} 

\begin{document}

\title {Anderson localization of a Bose-Einstein condensate in a 3D random potential}

\author{S.E. Skipetrov}

\author{A. Minguzzi}

\author{B.A. van Tiggelen}

\affiliation{Universit\'e Joseph Fourier, Laboratoire de Physique et Mod\'elisation des Mileux Condens\'es, CNRS, 25 Rue des Martyrs, BP 166, 38042 Grenoble, France}

\author{B. Shapiro}

\affiliation{Universit\'e Joseph Fourier, Laboratoire de Physique et Mod\'elisation des Mileux Condens\'es, CNRS, 25 Rue des Martyrs, BP 166, 38042 Grenoble, France}

\affiliation{Department of Physics, Technion-Israel Institute of Technology,
Haifa 32000, Israel}

\date{\today}

\begin{abstract}
We study the effect of Anderson localization on the expansion of a Bose-Einstein condensate, released from a harmonic trap, in a 3D random potential. We use scaling arguments and the self-consistent theory of localization to show that the long-time behavior of the condensate density is controlled by a single parameter equal to the ratio of the mobility edge and the chemical potential of the condensate. We find that the two critical exponents of the localization transition determine the evolution of the condensate density in time and space.
\end{abstract}
\maketitle
Anderson localization  \cite{anderson58} is an ubiquitous phenomenon which occurs in the propagation of waves  (electrons in solids \cite{anderson58}, microwaves \cite{chabanov00}, light \cite{wiersma97}, ultrasound \cite{weaver90}) in random media. Multiple scattering from
random obstacles  and the resulting destructive interference
suppress  wave propagation and can lead to exponentially localized eigenstates.

Recently there has been much interest in the possibility of
observing Anderson localization of Bose-Einstein condensates  obtained by trapping and cooling bosonic atoms \cite{lye05, sanchez07, shapiro07, shklovskii07}.
 A Bose-Einstein condensate is characterized by a macroscopic occupation of a single quantum state \cite{pitaevski03} and hence exhibits quantum, wave-like behavior despite its macroscopic size. Atomic Bose-Einstein condensates subjected to random external (optical) potentials are
potentially  good
candidates for observing Anderson localization of matter waves. Up to now,  the experiments have focused on 1D configurations \cite{lye05},  where all single-particle eigenstates are localized. In a typical experiment, the condensate is created in an optical or magneto-optical trap. The trap is then turned off and the condensate is allowed to expand.

In this Letter we study the expansion of the Bose-Einstein condensate in
a 3D random potential. Unlike in 1D, a critical energy (the mobility edge $\epsilon_c$) 
exists in 3D which
separates  extended and localized states. An
eigenstate is extended (localized) if
the corresponding energy is larger (smaller) than
$\epsilon_c$.
When the condensate is released from the trap,
the atoms achieve kinetic energies up to the chemical potential $\mu$ of the trapped condensate. 
For weak disorder $\epsilon_c < \mu$, and 
a fraction of atoms diffuses away, whereas the remainder
is localized, as was pointed out in \cite{shapiro07}.
We show here that, surprisingly, even for strong disorder $\epsilon_c > \mu$ only a fraction of the condensate will be localized. 
We study the full dynamics of the condensate expansion by accounting
for weak localization at energies $\epsilon > \epsilon_c$, strong localization at $\epsilon < \epsilon_c$, and critical behavior around the mobility edge.
Our main result is  that the effect of disorder on the expansion of the condensate is controlled
by a single parameter $\epsilon_c/\mu$, and that Anderson localization 
plays an important role even when the chemical potential of the condensate $\mu$ is much larger than the mobility edge $\epsilon_c$. We show that the behavior of
the average condensate density ${\bar n}({\bf r}, t)$ at large distances $r$
and long times $t$ is governed by the critical exponents $\nu$ and $s$ of the localization transition.
 This could provide a direct way to measure these exponents. The density of the localized part of the condensate ${\bar n}({\bf r}, \infty)$ does not decay exponentially with $r$, as one could have expected, but follows a power law.
 
 Consider a Bose-Einstein condensate of $N \gg 1$ atoms of mass $m$ trapped in a 3D spherically-symmetric harmonic potential $V_{\omega}({\bf r})$, characterized by the trap frequency $\omega$, to which we add a Gaussian uncorrelated random potential $V({\bf r})$:
 $B(\Delta {\bf r}) = \overline{V({\bf r}) V({\bf r} + \Delta {\bf r})} = u \delta(\Delta {\bf r})$, where the horizontal bar denotes averaging over an ensemble of realizations of the random potential $V({\bf r})$.
The wave function $\psi({\bf r}, t)$ of the condensate obeys the  (mean-field) Gross-Pitaevskii equation:
 \begin{eqnarray}
 i \hbar \frac{\partial \psi}{\partial t} &=&
 \left[ -\frac{\hbar^2}{2 m} \nabla^2 + V_{\omega}({\bf r}) + V({\bf r}) +
 g \left| \psi \right|^2 \right] \psi,
  \label{gpe}
 \end{eqnarray}
where $g$ measures the strength of repulsive interactions between atoms.
We consider a situation in which the random potential $V({\bf r})$ is initially absent \cite{sanchez07,shapiro07}. A sudden turn off of the confining potential $V_{\omega}({\bf r})$ then gives rise to a rapid expansion of the condensate. During a time $t \gtrsim t_0 = 1/\omega$ the initial potential energy of the trapped condensate is converted into kinetic energy  \cite{pitaevski03,castin96}.
The kinetic energy per atom becomes typically of the order of the chemical potential $\mu$ of the trapped condensate.
 The calculation that we present below applies both in the regime of strong ($\mu \gg \hbar \omega$) and weak ($\mu \sim \hbar \omega$) interactions. Whereas the former is realized in current experiments \cite{lye05}, the latter can, in principle, be reached using magnetic Feshbach resonances \cite{roati07}.

The random potential is switched on at a time $t > t_0$.  This creates a new energy scale at the mobility edge $\epsilon_c$.  At this stage the kinetic energy of the condensate becomes much larger than its interaction energy, and we can set $g = 0$ in Eq.\ (\ref{gpe}). The neglect of the nonlinear term in Eq.\ (\ref{gpe}) for times $t > t_0$ can be justified in the absence of disorder \cite{pitaevski03} and has been validated in the presence of a random potential by numerical simulations in 1D \cite{sanchez07}. Based on this, we expect that the neglect of interactions at long times $t > t_0$ should be valid in 3D as well, at least when the interactions are not too strong. The analysis of the complete nonlinear equation (\ref{gpe}) is a formidable task that falls far beyond the scope of this Letter. Even in 1D controversial numerical results have been reported recently \cite{kottos04}. We emphasize that even though we neglect interactions between atoms for $t > t_0$, we fully take them into account at earlier times $t < t_0$ and the strength of interactions $g$ enters our final results through $\mu$.
 
 The experimentally relevant quantity is the time-dependent condensate density
 $n({\bf r}, t) = | \psi({\bf r}, t) |^2$. For large distances $r \gg \ell$ (where $\ell = \hbar^4 \pi/u m^2$ is the mean free path  \cite{akkermans07}) and at long times $t \gg \hbar/\epsilon_c$, we derive the following expression for the condensate density, averaged over an ensemble of random realizations of $V({\bf r})$,
 \begin{eqnarray}
 {\bar n} ({\bf r}, t) &=&
 \int \frac{d^3 {\bf k}}{(2 \pi)^3}  \left| \phi({\bf k}) \right|^2
 \int_{-\infty}^{\infty} d \epsilon A({\bf k}, \epsilon)
 P_{\epsilon}({\bf r}, t).
 \label{averagen}
 \end{eqnarray}
 Here $P_{\epsilon}({\bf r}, t)$ is the probability density to find a particle of energy $\epsilon$, initially located at the origin, in the vicinity of ${\bf r}$ after a time $t$ (``probability of quantum diffusion'' \cite{akkermans07}), $A({\bf k}, \epsilon)$ is the spectral function, and $\phi({\bf k})$ is the Fourier transform of the wave function $\phi({\bf r})$ of the condensate after the first stage of expansion (i.e. after a time $t \gtrsim t_0$).
Physically, 
$d^3 {\bf k} | \phi({\bf k}) |^2/(2 \pi)^3$ is the number of atoms with a momentum around ${\hbar \bf k}$ at the time when the random potential is switched on, and
$d\epsilon A({\bf k}, \epsilon)$ is the probability to find an atom with energy around $\epsilon$ among all atoms with momentum $\hbar {\bf k}$, in the presence of the random potential.
In terms of the Green's function of the Schr\"{o}dinger equation  $G({\bf r}, t)$,
 the Fourier transform of $P_{\epsilon}({\bf r}, t)$
with respect to time $t$ is $P_{\epsilon}({\bf r}, \Omega) = \overline{G({\bf r}, \epsilon + \Omega/2)
G^*({\bf r}, \epsilon - \Omega/2)}/2 \pi \nu_{\epsilon}$, where
$\nu_{\epsilon}$ is the density of states, and
 $A({\bf k}, \epsilon) = -{\rm Im} {\bar G}({\bf k}, \epsilon)/\pi$. In previous work \cite{shapiro07}, a free-space expression $A({\bf k}, \epsilon) \propto \delta(\epsilon - \epsilon_{\bf k})$ was adopted with $\epsilon_{\bf k} = \hbar^2 {\bf k}^2/2m$. This appears to be a bad approximation for energies near and below the mobility edge, where the uncertainty in energy is large due to disorder.
 
To find the spectral function $A({\bf k}, \epsilon)$ we go beyond the first-order Born approximation \cite{sanchez07,kuhn07}, and use the so-called self-consistent Born approximation. We solve self-consistently the equations for the Dyson Green's function ${\bar G}$: ${\bar G}({\bf k}, \epsilon) = [\epsilon - \epsilon_{\bf k} - \Sigma({\bf k}, \epsilon)]^{-1}$, and the self-energy $\Sigma$:
$\Sigma({\bf k}, \epsilon) = \int d^3 {\bf k}^{\prime} B({\bf k} - {\bf k}^{\prime})
{\bar G}({\bf k}^{\prime}, \epsilon)$. For the uncorrelated random potential we obtain
 ${\bar G}({\bf k}, \epsilon) = 
 [ \epsilon - \epsilon^* - \epsilon_c/4 - \epsilon_{\bf k} + i  \hbar/2 \tau_{\epsilon}]^{-1}$  for $\epsilon > \epsilon^*$.
 The mean free time is
 $\tau_{\epsilon} = \pi \hbar^4/\sqrt{2} m^{3/2} u \sqrt{\epsilon - \epsilon^*}$.
 The edge of the spectrum $\epsilon^*$ depends on a cutoff needed to regularize the divergence of the integral over ${\bf k}^{\prime}$ in the equation for $\Sigma$. The value of $\epsilon^*$ is not important for the rest of our analysis because all relevant quantities depend on $\epsilon - \epsilon^*$.
 The mobility edge $\epsilon_c$ is assumed to obey the condition $k(\epsilon_c) \ell = 1$ and is located at $\epsilon_c = \epsilon^* + \hbar^2/2 m \ell^2$.   
 
 Another important ingredient of Eq.\ (\ref{averagen}) is the probability of quantum diffusion $P_{\epsilon}({\bf r}, t)$. At large $r \gg \ell$ and $\Omega \ll \epsilon$ its Fourier transform can be found in the hydrodynamic limit of quantum transport theory: $P_{\epsilon}({\bf r}, \Omega) = \exp[-r \sqrt{-i \Omega/D_{\epsilon}(\Omega)}]/4 \pi D_{\epsilon}(\Omega) r$ \cite{lee85}, where $D_{\epsilon}(\Omega)$ is the dynamic diffusion coefficient.
 As we will see below, the large-$r$ behavior of ${\bar n}({\bf r}, t)$ is dominated by atoms with energies $\epsilon \gtrsim \epsilon_c$, which justifies the use of the above expression for $P_{\epsilon}({\bf r}, \Omega)$ for $t \gg \hbar/\epsilon_c$. 

 Finally, the momentum distribution of the expanded condensate at the time when the random potential is switched on is assumed to be given by
 $| \phi({\bf k}) |^2 \propto 1 - k^2/k_{\mu}^2$ for $k < k_{\mu} = \sqrt{2 m \mu}/\hbar$ and $0$ otherwise. This expression follows from the dynamic scaling \cite{castin96} for $t \gg t_0$ and $\mu \gg \hbar \omega$. In the absence of interactions $\mu \sim \hbar \omega$ is a small energy scale.  If $\mu < \epsilon_c$, Eq.\ (\ref{averagen}) reduces to
${\bar n} ({\bf r}, t) = N \int_{-\infty}^{\infty} d \epsilon A({\bf 0}, \epsilon)
 P_{\epsilon}({\bf r}, t)$ and the precise profile of $| \phi({\bf k}) |^2$ has no importance.

In the following we will discriminate between localized ($\epsilon < \epsilon_c$), diffusing ($\epsilon > \epsilon_c$), and anomalously diffusing  ($\epsilon \simeq \epsilon_c$) atoms.
In the limit of very long times atoms with energies above the mobility edge have diffused away, and only localized atoms contribute to the condensate density ${\bar n}({\bf r}, t \rightarrow \infty)$ at any finite distance $r$. For these atoms we can set $D_{\epsilon}(\Omega) = -i \Omega \xi^2(\epsilon)$ \cite{lee85,abrahams79}. Inspection of Eq.\ (\ref{averagen}) then reveals that ${\bar n}({\bf r}, \infty)$ is determined by the critical dependence of the localization length $\xi$ on the energy $\epsilon$ in the vicinity of the mobility edge $\epsilon_c$: $\xi(\epsilon) \propto |\epsilon - \epsilon_c|^{-\nu}$. We obtain
\begin{eqnarray}
{\bar n}({\bf r}, \infty) \propto f \left( \epsilon_c/\mu \right)
\frac{N}{r^3} \times \left( \frac{\ell}{r} \right)^{1/\nu},
\label{loc}
\end{eqnarray}
where $f(x) \propto x^{3/2}$ for $x \ll 1$ and $f(x) \simeq \mathrm{const}$ for $x \gg 1$.
This important result demonstrates that the critical exponent $\nu$ can be determined from the spatial profile of the condensate density at long times. While the value of $\nu$ is not known for continuous disordered potentials, numerical solutions of the Anderson tight-binding model yield $\nu \simeq 1.5$ \cite{mackinnon83}, the self-consistent theory of localization predicts $\nu = 1$ \cite{vollhardt80}, and a claim of $\nu \simeq 0.5$ has been recently made for light in TiO$_2$ powders \cite{aegerter06}. 
  
The stationary density profile (\ref{loc}) may take a long time to be reached, especially for large distances $r$. It is therefore important to look for dynamical signatures of Anderson localization. The complete density of the condensate can be represented as a sum of the stationary part, considered above, and a time-dependent part: ${\bar n}({\bf r}, t) = {\bar n}({\bf r}, \infty) + \delta {\bar n}({\bf r}, t)$. At a given $r \gg \ell$ and for short times, $\delta {\bar n}({\bf r}, t)$ is dominated by the fastest atoms of the condensate. For weak disorder ($\epsilon_c \ll \mu$), these atoms typically have kinetic energies of order $\mu$, well above the mobility edge. Hence, they are almost unaffected by localization effects and diffuse with the ``classical'' diffusion coefficient $D_{\epsilon}(\Omega) \simeq D_{\epsilon}^{(0)} = \ell^2/3 \tau_{\epsilon}$. The integral over energies in Eq.\ (\ref{averagen}) can be then evaluated using the saddle point method.
For a given distance $r$, Eq.\ (\ref{averagen}) reaches a maximum at the ``arrival time''
\begin{eqnarray}
t_{\mathrm{arrival}} \simeq \frac{r^2}{6 D_{\mu}^{(0)}}.
\label{tarrival}
\end{eqnarray}
This result is equal to that found for a quasi-monochromatic wave packet with central energy $\mu$. 

At long times and for any $\epsilon_c/\mu$, the dynamic part of the condensate density is dominated by the critical behavior of $D_{\epsilon}(\Omega)$ at energies $\epsilon \gtrsim \epsilon_c$: $D_{\epsilon}(\Omega)  \propto | \epsilon - \epsilon_c|^s$ \cite{lee85,abrahams79}. The contribution of the critically diffusing atoms to the atomic density is
\begin{eqnarray}
\delta {\bar n}({\bf r}, t) 
\propto f(\epsilon_c/\mu)
\frac{N}{r^3} \left( \frac{r^2}{D_{\epsilon_c}^{(0)} t} \right)^{1/s},
\label{cd}
\end{eqnarray}
where the function $f(x)$ is the same as in Eq.\ (\ref{loc}) and where $s > 2/3$ is required to assure convergence of the integration over energies.
 Equation (\ref{cd}) reveals that the critical exponent $s$ can be directly measured in an experiment by observing the dynamics of expansion of the atomic cloud at large times.
The scaling theory of localization predicts $s = \nu$ for 3D disorder \cite{abrahams79}. A measurement of the critical exponents in an experiment with a Bose-Einstein condensate could provide a spectacular test of the (one-parameter) scaling theory.
  Note that the decay of  $\delta {\bar n}$ with time predicted by Eq.\ (\ref{cd}) is slower than what would have been obtained if localization were neglected ($\delta {\bar n} \sim 1/t^{3/2}$). Hence, localization effects modify the dynamics of the condensate expansion even when $\epsilon_c \ll \mu$.
 
An important time scale can be obtained from the comparison of Eqs.\ (\ref{loc}) and (\ref{cd}). Indeed, the convergence of the time-dependent profile 
${\bar n}({\bf r}, t)$ to ${\bar n}({\bf r}, \infty)$ given by Eq. (\ref{loc}) should become apparent when Eqs.\ (\ref{loc}) and (\ref{cd}) become of the same order. This happens at the ``localization time''  
\begin{eqnarray}
t_{\mathrm{loc}} \propto \frac{\hbar}{\epsilon_c} \left( \frac{r}{\ell} \right)^{2+s/\nu}.
\label{tloc}
\end{eqnarray} 
If we accept that $\nu = s$ in 3D \cite{abrahams79}, we obtain $t_{\mathrm{loc}} \propto r^3$ independent of $\nu = s$.
Note that the localization time (\ref{tloc}) exceeds the arrival time (\ref{tarrival}), if $r$ is larger than the healing length $1/k_{\mu}$
of the initial condensate.

We now consider the anomalously diffusing atoms with energies very close to the mobility edge. Indeed, in a narrow energy strip
$| \epsilon - \epsilon_c | < \epsilon_c (|\Omega|/\epsilon_c)^{1/3s}$ 
the critical diffusion, that led to Eq.\ (\ref{cd}), is taken over by the anomalous diffusion:
$D_{\epsilon}(\Omega) \propto (-i \Omega)^{1/3}$ \cite{wegner76}. Analysis shows that for $t > t_{\mathrm{loc}}$ the anomalously diffusing atoms dominate the dynamic part of the condensate density: 
\begin{eqnarray}
\delta {\bar n}({\bf r}, t) 
\propto f(\epsilon_c/\mu)
\frac{N}{r \ell^2} \left( \frac{\epsilon_c t}{\hbar} \right)^{-2/3 - 1/3s}.
\label{ad}
\end{eqnarray} 
However, at $t \sim t_{\mathrm{loc}}$ the density of anomalously diffusing atoms (\ref{ad}) is already of the same order as the density of localized atoms (\ref{loc}). Therefore,  the contribution of anomalously diffusing atoms may be difficult to observe in an experiment.   

\begin{figure}
\centering
\includegraphics[width=8cm,angle=0]{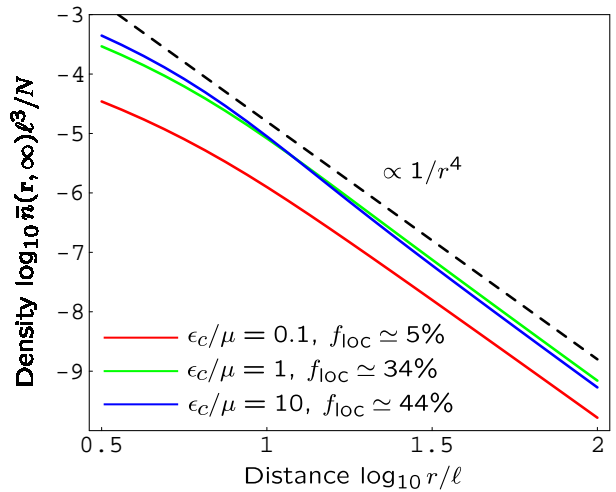} 
\caption{Profiles of the average atomic density, associated with the localized part of the atomic cloud, at long times. The solid lines are obtained from the self-consistent theory of localization for three different ratios of mobility edge $\epsilon_c$ and chemical potential $\mu$ of the initial condensate. For each $\epsilon_c/\mu$, the fraction of localized atoms $f_{\mathrm{loc}}$ is given. The dashed line shows the asymptote ${\bar n} \propto 1/r^4$ followed by all curves for large $r$.}
\label{fig1}
\end{figure}

\begin{figure}
\centering
\includegraphics[width=8cm,angle=0]{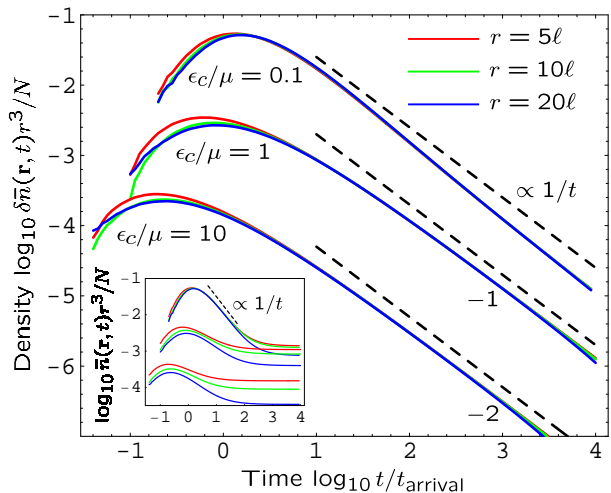}
\caption{Profiles of the dynamic part of the average atomic density obtained from the self-consistent theory of localization (solid lines). The time is given in units of $t_{\mathrm{arrival}}$ defined by Eq.\ (\ref{tarrival}). The dashed lines show $1/t$ asymptotes. At a given ratio of mobility edge $\epsilon_c$ and chemical potential $\mu$, $\delta {\bar n}({\bf r}, t) r^3/N$ for different $r$ (different colors) fall on a universal curve. For clarity, the curves corresponding to $\epsilon_c/\mu = 1$ and $10$ are shifted downwards by 1 and 2 units, respectively. The inset shows the complete atomic densities ${\bar n}({\bf r}, t)$ obtained by adding $\delta {\bar n}$ shown in the main plot and  ${\bar n}({\bf r}, \infty)$ of Fig.\ \ref{fig1}.}
\label{fig2}
\end{figure}

Up to here we have presented a general analysis based on scaling ideas only. The self-consistent theory of localization \cite{vollhardt80} allows us to calculate ${\bar n}({\bf r}, t)$ from Eq.\ (\ref{averagen}) without any additional assumptions. In this theory $D_{\epsilon}(\Omega)$ obeys
\begin{eqnarray}
\frac{D_{\epsilon}^{(0)}}{D_{\epsilon}(\Omega)} =
1 + \frac{1}{\pi \nu_{\epsilon}}
\int_0^{q_{\mathrm{max}}} \frac{d^3 {\bf q}}{(2 \pi)^3}
P_{\epsilon}({\bf q}, \Omega),
\label{selfcon}
\end{eqnarray}
where $q_{\mathrm{max}} = \pi/3 \ell$.
The self-consistent theory predicts the critical exponents $\nu = s = 1$.
We plot ${\bar n}({\bf r}, \infty)$ and $\delta {\bar n}({\bf r}, t)$ obtained from Eqs.\ (\ref{averagen}) and (\ref{selfcon}) in Figs.\ \ref{fig1} and \ref{fig2}. 
As follows from Fig.\ \ref{fig1}, for sufficiently large $r$ the density profile ${\bar n}({\bf r}, \infty)$ decays as a power-law $1/r^4$ for all $\epsilon_c/\mu$, as predicted by Eq.\ (\ref{loc}) with $\nu = 1$. The dynamic part of the atomic density $\delta {\bar n}({\bf r}, t)$ is shown in Fig.\ \ref{fig2} and reaches a maximum at times of order of $t_{\mathrm{arrival}}$ given by Eq.\ (\ref{tarrival}), at least when $\epsilon_c \lesssim \mu$. The maximum shifts to shorter times when $\epsilon_c/\mu$ increases. As predicted by Eq.\ (\ref{cd}), the plots of $\delta {\bar n}({\bf r}, t) r^3/N$ for different $r$ fall on a universal curve,
when shown as functions of $t/t_{\mathrm{arrival}}$ at  fixed $\epsilon_c/\mu$.
All curves in Fig.\ \ref{fig2} follow the $1/t$ asymptote for long times. This is in agreement with Eq.\ (\ref{cd}) for $t < t_{\mathrm{loc}}$ and Eq.\ (\ref{ad}) for $t > t_{\mathrm{loc}}$, with $s = 1$.

The number of atoms that stay localized after the time $t_{\mathrm{loc}}$ can be found by integrating Eq.\ (\ref{averagen}) over ${\bf r}$ and by restricting the integration over energies to $\epsilon < \epsilon_c$. We find for the fraction of localized atoms $f_{\mathrm{loc}} \simeq (5/2)(\epsilon_c/\mu)^{3/2}$ for $\epsilon_c \ll \mu$ and $f_{\mathrm{loc}} \simeq 0.45$ for $\epsilon_c \gg \mu$. 
Remarkably, this result is independent of the model used for the localization length $\xi(\epsilon)$ and, in particular, independent of the critical exponent $\nu$. 
The conclusion that even for strong disorder ($\epsilon_c \gg \mu$) only a fraction of the condensate stays localized is a surprising outcome. 
It is due to the large-energy tail of the spectral function  derived above for the uncorrelated disorder: $A({\bf k}, \epsilon) \propto 1/\epsilon^{3/2}$. Atoms with high energies exhibit diffuse behavior, even though their initial kinetic energies were small ($\epsilon_{\bf k} \ll \mu$). This explains the small value of $f_{\mathrm{loc}}$. It is straightforward to show that for a potential with correlation radius $r_0$ the large-energy tail of $A({\bf k}, \epsilon)$ is suppressed for energies larger than $\epsilon_0 \sim \epsilon_c (\ell/r_0)^2$. Thus,  correlations seem to help in localizing more atoms.

The available 1D experiments \cite{lye05} and theoretical developments \cite{kuhn07} allow us to estimate the minimal realistic values of the mean free path $\ell$ in 3D optical speckle potentials to be in the range of 1 to 10 $\mu$m. From this values we obtain $\epsilon_c/\mu \sim 10^{-4}$--$10^{-2}$ for $^{87}$Rb atoms and $\mu/\hbar \sim 10$ kHz \cite{lye05}. The corresponding arrival (localization) time is in the range of 1 to 10 ms (100 ms to 100 s) for $r = 5 \ell$. With the lifetimes of Bose-Einstein condensates attaining 20 s \cite{lean03}, these estimations demonstrate that our theoretical predictions are within the reach of current experiments.

In conclusion, we have shown that the long-time large-distance behavior of the average density ${\bar n}({\bf r}, t)$ of a Bose-Einstein condensate expanding in a 3D random potential is governed by a single parameter equal to the ratio of the mobility edge $\epsilon_c$ to the chemical potential $\mu$ of the condensate. The critical exponents of the Anderson localization transition $\nu$ and $s$ determine the evolution of ${\bar n}({\bf r}, t)$ in time and space. Our results open a new way to measure the critical exponents in an experiment.

\begin{acknowledgments}
We thank Ch. Miniatura for many stimulating discussions.
SES acknowledges financial support of the French ANR  (project 06-BLAN-0096 CAROL) and the French Ministry of Education and Research.
AM benefited from discussions with B. DeMarco and M. Inguscio, and from a CNRS-UIUC exchange.

\end{acknowledgments}

\end{document}